\documentclass[dvips]{article}
\usepackage{icrctc07}

\def\la{\lower.5ex\hbox{$\; \buildrel < \over \sim \;$}}
\def\ga{\lower.5ex\hbox{$\; \buildrel > \over \sim \;$}}

\title{Propagation of UHE protons through a magnetized large scale structure}
\shorttitle{UHECRs}

\authors{Hyesung Kang$^{1}$, Santabrata Das$^{2}$, Dongsu Ryu$^{3}$,
Jungyeon Cho$^{3}$ }
\shortauthors{Hyesung Kang and et al.}

\afiliations{
$^1$Department of Earth Sciences, Pusan National University, Pusan 609-735,
Korea:\\ 
$^{2}$ARCSEC, Sejong University, Seoul 143-747,
Korea:\\ 
$^{3}$Department of Astronomy \& Space Science, Chungnam National University,
Daejeon 305-764, Korea:\\}
\email{kang@uju.es.pusan.ac.kr, sbdas@canopus.cnu.ac.kr, ryu@canopus.cnu.ac.kr,
cho@canopus.cnu.ac.kr}

\abstract{
The propagation of UHECRs is affected by the intergalactic magnetic
fields that were produced during the course of the large scale structure
formation of the universe.  We adopt a novel model where
the large scale extragalactic magnetic fields (EGMF) are estimated
from local dynamic properties of the gas flows in hydrodynamic
simulations of a concordance $\Lambda$CDM universe. With
the model magnetic fields, we calculate the deflection angle, 
time delay and energy spectrum of protons with $E > 10^{19}$ eV 
that are injected at cosmological sources and then travel through 
the large scale structure of the universe, losing the energy due 
to interactions with the cosmic background radiation.  
Implications of this study on the origin of UHECRs are discussed. 
}

\begin{document}
\maketitle

\section{Introduction}

The enigma of ultrahigh energy cosmic rays (UHECRs) with energy
$E \ga$ 5 EeV (1 EeV = $10^{18}$ eV) has received considerable attention, 
both observationally and theoretically, over the past decades \cite{cronin99,
nagano00}. 
Although the composition studies indicate that most of them are likely to be protons 
\cite{abbasi05}, the nature and origin of these CRs remain yet to be understood.
At these high energies, UHECRs are not confined within
the galactic magnetic plane and therefore their origin should be
extragalactic. In particular, the isotropic distribution of arrival 
directions suggests that large number of UHECR sources might be 
distributed at sufficiently large distances \cite{nagano00,bo03}.
UHECRs from cosmological sources get significantly attenuated 
during their propagation due to interaction with 
the cosmic microwave background, 
resulting in the so-called Greisen-Zatsepin-Kuzmin (GZK) cutoff 
in the energy spectrum \cite{bgg06}. 
Interestingly, AGASA experiment failed to identify this cutoff \cite{nagano00}, 
whereas HiRes recorded it \cite{z04}. 
Detection of super-GZK events with isotropic arrival directions is
enigmatic, because powerful astronomical objects are very rare in the local universe. 
New results from Pierre Auger experiment reported in this conference
may shed light on this puzzle.

Charged UHECRs are expected to be deflected by large scale extragalactic 
magnetic fields (EGMF), so their arrival directions can deviate from 
source directions and arrival time is delayed compared to the light
travel time \cite{armen05,detal05}. 
At present, the main challenge is to estimate the EGMF,
since observing it is still a difficult task. From
this standpoint, we employ a novel model for the EGMF
in which the strength of the intergalactic magnetic fields is estimated
from local turbulence in a cosmological hydrodynamic
simulation. 
The trajectories of UHE protons from active galactic nuclei (AGNs),
propagating through the model EGMF, are calculated, 
including all relevant energy loss processes. 
We then study the angular deflection, time delay 
and energy spectrum of UHE protons that arrive in the regions
similar to the Local Group.  

\begin{figure*}[th]
\begin{center}
\vskip -4.1cm
\includegraphics[width=0.8\textwidth]{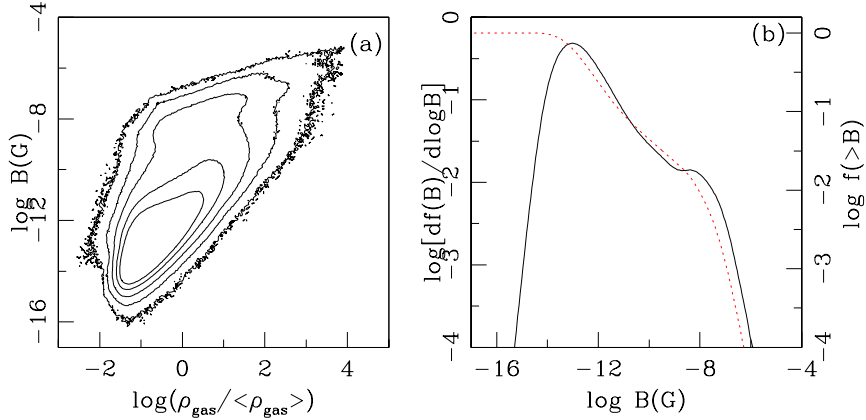}
\end{center}
\vskip -3.3cm
\caption{
(a) Volume fraction in the gas density-EGMF strength plane with our model EGMF
at $z = 0$.
(b) Volume fraction, $df/d \log(B)$, (solid line) 
and its cumulative distribution, $f(>B)$, (dotted)
as a function of EGMF strength.}
\label{fig1}
\end{figure*}

\section{Model}

\subsection{Extragalactic Magnetic Fields}

Concordance $\Lambda$CDM cosmological simulations are carried out along 
with passively evolved magnetic fields in a cubic box of comoving size 
$100 h^{-1}$ Mpc, using $512^3$ grid zones with the following
parameters: $\Omega_{BM}=0.043$,
$\Omega_{DM}=0.227$, and $\Omega_{\Lambda}=0.73$,
$h \equiv H_0$/(100 km/s/Mpc) = 0.7, and $\sigma_8 = 0.8$.
Six different initial realizations are used.
Only the directional information of the passive fields is 
adopted.
Assuming that the intergalactic magnetic fields result from
turbulent motion of the intergalactic gas, the strength of the EGMF is computed 
directly by equating the magnetic energy to a suitable fraction 
of the turbulent energy of the intergalactic gas
(see \cite{rk07} for details).

Fig. 1 shows the volume fraction in the $\rho_{gas}$ - B plane in our model. 
The EGMF are correlated with the large scale structure: 
the strongest around clusters of galaxies and the weakest in voids.
In the regions of galaxy clusters with 
$\rho_{gas}/\langle \rho_{gas} \rangle \ga 10^3$, 
$\langle B \rangle \sim 10^{-6}$G.
For typical filamentary regions with $\rho_{gas}/\langle \rho_{gas} \rangle
\sim 30$, $\langle B \rangle \sim 10^{-8}$G, while
in void region, $\langle B \rangle \sim 10^{-12}$G.
 
\begin{figure*}[th]
\begin{center}
\vskip -4.cm
\includegraphics[width=0.8\textwidth]{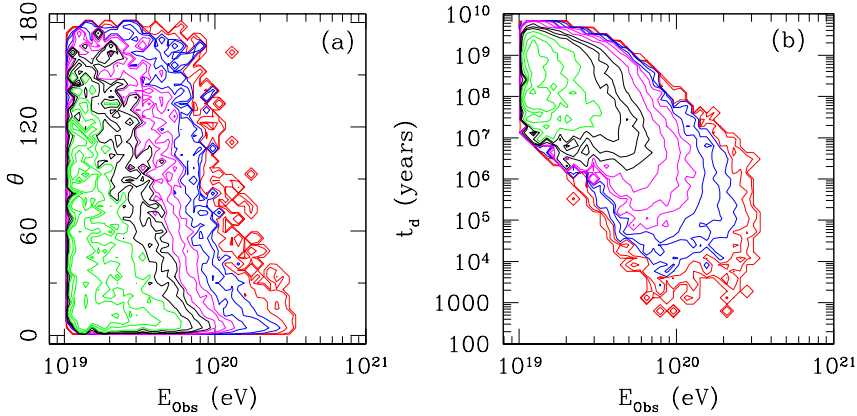}
\end{center}
\vskip -3.5cm
\caption{
(a) Distribution of events in the plane of observed particle
energy and deflection angle.
(b) Distribution of events in the plane of observed particle
energy and time delay.
Protons are injected with a power-law spectrum of $\gamma=2.7$ at sources.
}\label{fig2}
\end{figure*}

\subsection{CR Sources and Observers}

Assuming that UHECRs originate from AGNs,
we identify galaxy clusters with $kT>1.0$ keV in the cosmological simulation
as sources, since AGNs 
are likely to form in high density environment.
There are 20-30 such clusters in the simulated volume, 
corresponding to the source density of $n_s = 2-3 \times10^{-5} 
h^{3}{\rm Mpc}^{-3}$ and the mean separation of $l_s \sim 40 h^{-1}{\rm Mpc}$.
The field strength in source locations ranges 
$ 0.1 <B_s(\mu {\rm G})< 10$ with $\langle B_s \rangle \sim 1\mu$G.
UHE protons with a power-law energy spectrum of $E^{-\gamma}$ 
for $10\ {\rm EeV} \le E \le 10^3 {\rm EeV}$
are distributed among the sources and then launched in random directions.
We follow the trajectory of individual protons by numerically integrating
the equation of motion in the model EGMF, considering continuous energy loss
due to photo-pair and photo-pion production \cite{bgg06}. 

An observer is placed at each $10^3$ groups of galaxies with
$ 0.05 {\rm keV} < kT < 0.5 {\rm keV}$ in order to
select locations similar to the Local Group.
These groups are not distributed uniformly, but located mostly along
filaments, following the matter distribution of the large scale structure.  
Around observer locations, 
$ 10^{-4} <B_{obs}({\mu \rm G})< 0.1$ with $\langle B_{obs} \rangle \sim 10$nG.
Once a particle visits an observer within a sphere of $R_{obs}=0.5 h^{-1}$ Mpc,
the arrival direction, time delay and energy of the particle are
registered as an `observed' event. 
We let the particle continue its journey, visiting
several groups during its full flight,  
until its energy is reduced to 10 EeV.
With $10^4$ protons injected at sources, 
about $10^5$ events are recorded.

\section{Results}

\subsection{Deflection Angle and Time Delay}

Note that the gyroradius of proton is 
$r_g = 10{\rm kpc}\ (E/10^{19}{\rm eV}) (B/\mu{\rm G})^{-1}$.
With our model EGMF, the trajectories of $\sim$10 EeV protons
could be severely deflected when they leave the source
regions with $\langle B_s \rangle \sim 1\mu$G
and when they fly by the cluster/group regions with $B \ga 10^{-8}$G, 
since clusters/groups
have a typical radius of $\langle R \rangle \sim 1 {\rm Mpc}$.
According to Fig. 1 the volume filling factor 
with $B >10^{-8}{\rm G}$ is 
$f(>10^{-8}{\rm G}) \approx 0.02$. 

In Fig. 2(a) we present the distribution of the deflection angle between
the arrival direction of UHE protons and the source direction, $\theta$, as
a function of the observed energy.
This distribution depends rather sensitively on the power-law index of the injection
spectrum, $\gamma$, and the minimim distance between sources and 
observers, $D$.
Here, the results of the case with $\gamma=2.7$ and $D=20$Mpc are
presented.
On average the deflection angle decreases with the energy, 
showing a transition from diffuse transport regime to 
rectilinear propagation region at $E\sim$ 100 EeV.
For $E\ge40$ EeV  30 \% of `observed' events arrive with $\theta \le 5^{\circ}$
from source directions,
while 60 \% with $\theta \le 5^{\circ}$ for $E\ge100$ EeV. 
Some super-GZK protons can avoid flying through strong field regions
around clusters and do not get significantly deflected by weak fields in filamentary
structures.   

Since the deflected path length is longer than the rectilinear distance,
the particle travel time is delayed compared to the light travel time.
In Fig. 2(b), we show the distribution of $t_d$,
the delay of the particle travel time compared to the light travel
time, as a function of the observed energy.
The rectilinear travel time for the mean separation of sources, 
$t_{rec}\equiv l_s/c \approx 10^{8.25}$ years.
On average the time delays are longer for lower energy particles, with
$\langle t_d\rangle \sim 10^{9} {\rm yrs} \sim 6 t_{rec}$ around 10 EeV and
$\langle t_d\rangle \sim 10^{5} {\rm yrs} \sim 10^{-3} t_{rec}$ above 100 EeV. 
\subsection{Energy Spectrum}

Fig. 3 shows the predicted energy spectra of UHE proton events
for the injection spectrum of $N_{inj}(E)\propto E^{-\gamma}$ with
$\gamma = 2.0$, 2.4 and $2.7$ and $D=5,$ 10, and 20 Mpc.
We note that $J(E)E^3$ in the plot is given in an arbitrary
units, since the amplitude of the injected spectrum at sources
is arbitrary.
In all the cases the presence of GZK supression above 50 EeV is
obvious, so the AGASA data cannot be consistent with our results.
In case of $D=5$ Mpc, there is a significant number of particles above 100 EeV
that come mainly from nearby sources. 
Below 50 EeV, the spectrum gets flatter due to the pileup of the
particles that had higher initial energies but have lost their energy 
via photo-pion production.
The injection spectrum with $\gamma=2.7$ seems to produce the best
fit for the HiRes-2 data, although the case with $\gamma=2.4$ can 
be considered as being consistent with the HiRes observations. 

\begin{figure}[th]
\begin{center}
\vskip -0.5cm
\includegraphics [width=0.5\textwidth]{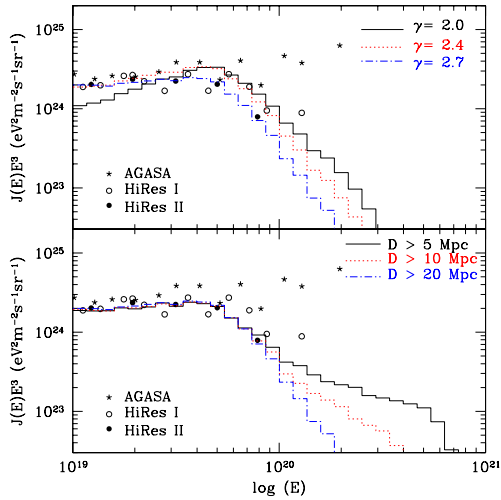}
\end{center}
\vskip -1.0cm
\caption{Predicted energy spectra of UHE protons
that are injected with
a power-law spectrum of $N_{inj}(E)\propto E^{-\gamma}$ at sources
and propagated 
through a universe with the adopted model EGMF. 
{\it Top panel:} the case with the minimum source-observer distance $D=$ 20Mpc
is shown for $\gamma = 2.0$ (solid line),
2.4 (dotted) and 2.7 (dot-dashed).
{\it Bottom panel:} the case with $\gamma=2.7$ is shown for 
$D = 5$ Mpc (solid line),
10 Mpc (dotted) and 20 Mpc (dot-dashed).
Filled circles are the observational data from HiRes-2 \cite{z04}.
and open circles are from HiRes-1, 
while stars are from AGASA \cite{nagano00}.
}\label{fig3}
\end{figure}

\section{Conclusion}

We study the propagation of UHE protons that originate from
cosmological sources located at galaxy clusters, travel through 
the EGMF correlated with the large scale structure of the universe,
and arrive in the regions similar to the Local Group.
A novel model for the EGMF based on turbulence
dynamo is adopted.
Since the volume filling factor, $f(>10^{-7}{\rm G})$ is less 
than $5\times 10^{-3}$,
about 60 \% of UHE protons above 100 EeV could avoid visiting strong 
field regions and arrive at Earth
within $5^{\circ}$ of their sources, although the deflection angle 
is overall distributed over a wide range. 
Below the GZK energy, the propagation is in the diffusion limit and
both the deflection angle and time delay are substantial.  
This indicates that in the present scenario, UHE proton astronomy
may be possible only for $E>$ 100 EeV. 
The predicted energy spectrum of UHE protons
exhibits the GZK supression, and fits the HiRes observations,
if the injection spectrum has $\gamma=2.4-2.7$.

\section{Acknowledgments}
HK and SD was supported by KOSEF through Astrophysical
Research Center for the Structure and Evolution of the Cosmos(ARCSEC).
DR was supported by the Korea Research Foundation Grant
funded by the Korean Government (MOEHRD) (KRF-2004-015-C00213).

\bibliography{icrc0418}
%This in the bibtex style, is ok.
\bibliographystyle{plain}

\end{document}